# Maximal Entropy Reduction Algorithm for SAR ADC Clock Compression

*Arkady Molev-Shteiman and Xiao-Feng Qi*

Radio Algorithms Research, Futurewei Technologies, New Jersey Research Center, Bridgewater, NJ

**ABSTRACT**

Reduction of comparison cycles leads to power savings of a successive-approximation-register (SAR) analog-to-digital converter (ADC). We establish that the lowest average number of comparison cycles of a SAR ADC approaches the entropy of the ADC output, and proposed a simple adaptive algorithm that approaches this lower bound. Today's SAR ADC uses binary search, which consumes more power than necessary for non-uniform input distributions commonly found in practice. We refer to a SAR ADC employing such a search algorithm the maximal entropy reduction ADC (MER ADC).

## 1. INTRODUCTION

ADC is ubiquitous in every application that requires conversion of analog signals to digital form. Due to its relatively low cost and power consumption, the SAR ADC architecture sees wide-spread applications where conversion speed is not extremely high. However, minimizing SAR ADC power consumption is still critical for many applications that requires conversion of large amount of data. Examples include sensing, image processing and massive MIMO communication. The power consumption of SAR ADC is strongly dependent on the number of comparison cycles (clocks) required for each conversion. Conventional SAR ADC uses binary search algorithm with fixed number of cycles, which is equal to the number of ADC bits. However, it is optimal only for uniform input probability distribution. For other inputs, optimal strategy is to reduce the search length for high likelihood samples, to more than compensate for the increase in the search length for low likelihood samples. The average number of comparison cycles for such ADC decreases as a result, which means lower power consumption.

The SAR ADC with variable number of comparison cycles was proposed recently in [1] and [2], where a few algorithms of non-binary search were proposed for specific types of inputs.

Our work, on the contrary, generalizes to analysis of search algorithms from information theory point of view. We apply the method of Maximum Entropy Reduction (MER) decision tree ([3] and references therein) to the development of a generic search algorithm, which can be matched to any ADC output probability distribution. When ADC resolution approaches infinity, the average search length (number of comparison cycles consumed to determine the correct ADC output value) approaches the ADC output entropy. It matches theoretical lower bound for average decision cycles of binary decision tree presented in [3]. To our knowledge this is the first contribution applying decision tree bounds to comparison cycles of SAR ADC.

We also propose a simple adaptive search algorithm that allows efficient hardware implementation. It estimates the change in ADC output probability distribution and adjusts the decision thresholds accordingly.

Another method that benefit from non-uniform distribution of ADC input is non-linear quantization [4] [5]. It reduces the average quantization noise, by allocating low quantization steps in high likelihood regions and high quantization steps in low likelihood regions. However for many applications we cannot allow increase of quantization noise even in low likelihood region; Furthermore, if actual input distribution is different from the assumed one, ADC with non-linear quantization will suffer from performance degradation due to increase of average quantization noise. MER ADC is self-correcting: the number of comparison cycles increases momentarily, but settles to a new optimum through the ADC adaptation mechanism.

## 2. CONVENTIONAL SAR ADC SYSTEM MODEL

An ADC performs quantization $y = round(x/\Delta)$, where $x$ is ADC input, $y$ is ADC output, $round(\ )$ denotes rounding operation and $\Delta$ is the quantization step.

Assume that input to an *N*-bits ADC is limited by:
$$-0.5 \cdot \Delta \leq x < \left(2^N \cdot \Delta - 0.5 \cdot \Delta\right) \quad (1.1)$$

Then *y* is any integer number from 0 to $2^N - 1$ satisfying to:
$$\Delta \cdot (y - 0.5) \leq x < \Delta \cdot (y + 0.5) \quad (1.2)$$

The SAR ADC provides iterative search of output *y* satisfying (1.2). Its block diagram is shown in figure 1.

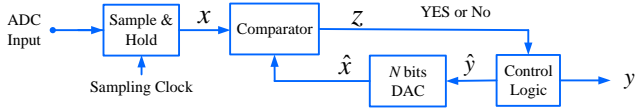

Figure 1  The block diagram of SAR-ADC

A SAR ADC contains the following components.

- A sample-and-hold unit that provides time domain sampling of input signal.
- An *N*-bits Digital-to-Analog Converter (DAC) that generates analog reference based on digital hypothesis $\hat{y}$
$$\hat{x}(\hat{y}) = \Delta \cdot (\hat{y} - 0.5) \quad (1.3)$$
- A comparator which compares input signal with DAC output and produces binary outcome:
$$z = (x < \hat{x}(\hat{y})) = (0 \text{ or } 1) \quad (1.4)$$
- Control logic that searches for the DAC input combination with guidance from comparator output, which satisfies $\hat{x}(\hat{y}) \leq x < \hat{x}(\hat{y}) + \Delta$. The combination is taken as the ADC output, namely $y = \hat{y}$.

Let $\hat{y}_m$ and $z_m$ denote DAC input and comparator output following comparison cycle *m*, respectively.

After each comparison cycle *m* we may conclude that ADC output satisfies $(LB_m \leq y < UB_m)$, where $LB_m$ and $UB_m$ denote lower and upper bound of ADC output ambiguity interval, respectively: $B_m = [LB_m, UB_m]$. Obviously, before the search starts. $B_0 = [0, 2^N]$.

Each comparison cycle changes ADC output ambiguity interval according to the rule below.
$$\begin{array}{ll} \text{If } (x < \hat{x}(\hat{y}_m)) & UB_m = \hat{y}_m \quad LB_m = LB_{m-1} \\ \text{Else} & LB_m = \hat{y}_m \quad UB_m = UB_{m-1} \end{array} \quad (1.5)$$

A rational search algorithm chooses new comparison threshold $\hat{y}_{m+1}$ within previous ambiguity interval. Therefore each comparison cycle reduces the size of ADC output ambiguity interval until eventually (say after *M* comparison cycles) it shrinks to $(UB_M - LB_M) = 1$, which means that there is no more ambiguity and we found the ADC output which is equal to $y = LB_M$. The above procedure is essentially a search over a decision tree.

The conventional SAR ADC does binary search. Starting from ambiguity interval $[0, 2^N]$, each comparison cycle it cuts ambiguity interval by half, by choosing hypothesis right at the center of possibilities interval (ensuring equal size of each half). After *N* comparison cycles it shrinks ambiguity interval to one, ending the search process. The flowchart of binary search algorithm is given in Figure 2:

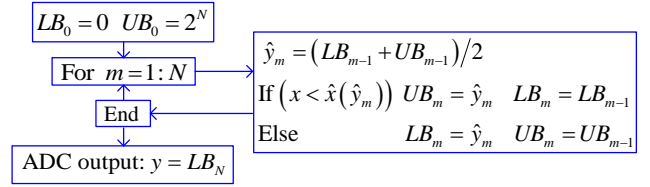

Figure 2  The flowchart of binary search algorithm

## 3. MAXIMUM ENTROPY REDUCTION SEARCH

An analytical lower bound on average decision cycles for a generalized decision tree was presented in [3]. In the case of SAR ADC, this limit corresponds to the entropy of the ADC output, which can simultaneously be considered as the initial value of the ADC output entropy before the search outcomes are available:
$$H(y) = -\sum_{n=0}^{2^N - 1} p_n \cdot \log_2(p_n) \quad (1.6)$$

where $p_n$ denote the probability of the ADC output *y* taking on the value *n*:
$$p_n = \Pr(y = n) = \Pr(-0.5 \cdot \Delta \leq (x - n \cdot \Delta) < 0.5 \cdot \Delta) \quad (1.7)$$

The *m*-th comparison changes ADC output ambiguity interval to $\bar{B}_m$, and the ADC output entropy after *m* comparison cycle conditioned on $B_m = \bar{B}_m$ becomes
$$H(y \mid B_m = \bar{B}_m) = -\sum_{n=\bar{LB}_m}^{\bar{UB}_m - 1} p_{m,n} \cdot \log_2(p_{m,n}) \quad (1.8)$$

where:
$$p_{m,n} = \Pr(y = n \mid B_m = \bar{B}_m) = p_n \bigg/ \sum_{n=\bar{LB}_m}^{\bar{UB}_m - 1} p_n \quad (1.9)$$

Search algorithm (1.5) ensures that each comparison cycle reduces ambiguity interval of ADC output, until it shrinks to one. When that happens the output entropy is equal to 0, and the search process ends. In order to minimize search length we must choose threshold $\hat{y}_{m+1}$ for the (*m*+1)-th comparison cycle in such a way that the post-comparison expectation of the ADC output entropy conditioned on the previous ambiguity interval $\bar{B}_m$ is minimal.

According to (1.5) ambiguity interval $B_{m+1}$ is fully determined by the previous ambiguity interval $B_m$ and comparison result $z_{m+1}$. It follows that ADC output entropy after ($m+1$) comparison cycle is:

$$H\left(y \mid B_{m+1} = \bar{B}_{m+1}\right) = H\left(y \mid B_m = \bar{B}_m, z_{m+1} = \bar{z}_{m+1}\right) \quad (1.10)$$

Therefore the expectation of the ADC output entropy after the ($m+1$)-th comparison given previous ambiguity interval $B_m$ is average entropy over two possible realizations of comparator output $z_{m+1}$:

$$E_m = \sum_{z=0,1} \Pr{}_m \left(z_{m+1} = z\right) \cdot H\left(y \mid B_m = \bar{B}_m, z_{m+1} = z\right) \quad (1.11)$$

which happens to be the conditional entropy of ADC output $y$ given comparison result $z_{m+1}$ conditioned on $\left(B_m = \bar{B}_m\right)$

$$E_m = H\left(y \mid z_{m+1}, B_m = \bar{B}_m\right) \quad (1.12)$$

According to chain rule [6]:

$$E_m = H\left(y, z_{m+1} \mid B_m = \bar{B}_m\right) - H\left(z_{m+1} \mid B_m = \bar{B}_m\right) \quad (1.13)$$

where $H\left(y, z_{m+1} \mid B_m = \bar{B}_m\right)$ is joint entropy of $y$ and $z_{m+1}$ conditioned on $\left(B_m = \bar{B}_m\right)$. Since the comparison result $z_{m+1}$ is completely determined by the ADC output $y$,

$$E_m = H\left(y \mid B_m = \bar{B}_m\right) - H\left(z_{m+1} \mid B_m = \bar{B}_m\right) \quad (1.14)$$

From (1.14) it follows that to minimize entropy expectation after comparison cycle ($m+1$) we must maximize entropy of comparator output in cycle ($m+1$), i.e.

$$\min(E_m) \Leftrightarrow \max\left(H\left(z_{m+1} \mid B_m = \bar{B}_m\right)\right) \quad (1.15)$$

Therefore we may also call $H\left(z_{m+1} \mid B_m = \bar{B}_m\right)$ the information gain (or entropy reduction) of comparison cycle ($m+1$). Because the comparator output $z_{m+1}$ is binary, its entropy is equal to:

$$H\left(z_{m+1} \mid B_m = \bar{B}_m\right) = -p_{z,0} \cdot \log_2(p_{z,0}) - p_{z,1} \cdot \log_2(p_{z,1}) \quad (1.16)$$

Where:

$$p_{z,0} = \Pr\left(z_{m+1} = 0 \mid B_m = \bar{B}_m\right) = \sum_{n=\bar{LB}_m}^{\hat{y}_{m+1}-1} p_n \Big/ \sum_{n=\bar{LB}_m}^{\bar{UB}_m-1} p_n \quad (1.17)$$

$$p_{z,1} = \Pr\left(z_{m+1} = 1 \mid B_m = \bar{B}_m\right) = \sum_{n=\hat{y}_{m+1}}^{\bar{UB}_m-1} p_n \Big/ \sum_{n=\bar{LB}_m}^{\bar{UB}_m-1} p_n \quad (1.18)$$

The entropy of binary output is maximized when the absolute value of the difference $|p_{z,0} - p_{z,1}|$ is minimized [6]. Therefore optimal threshold for comparison ($m+1$) is:

$$\hat{y}_{m+1} = \arg\min_{\hat{y}} \left(|p_{z,0} - p_{z,1}|\right) = \arg\min_{\hat{y}} \left(\left|\sum_{n=\bar{LB}_m}^{\hat{y}-1} p_n - \sum_{n=\hat{y}}^{\bar{UB}_m-1} p_n\right|\right) \quad (1.19)$$

As can be seen the thresholds and comparison outcomes are solely determined by the ADC outputs probabilities. The flowchart of MER search algorithm is shown in Figure 3.

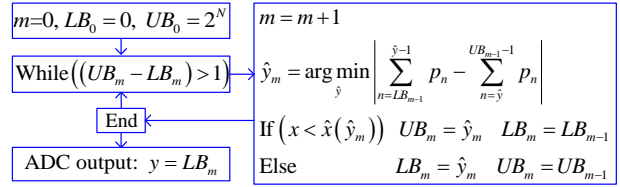

Figure 3 The flowchart of MER search algorithm

If ADC resolution $N$ is sufficiently high, each comparison splits the ADC output ambiguity intervals into two sub-intervals with almost equal probability and chooses one of them, and the probability of the subsequent ambiguity interval is decreased by a factor of 2 from the previous one. Suppose the input corresponds to ADC output $n$, and it takes $M$ comparison cycles for the MER ADC to locate this output. The probability of ambiguity interval after $M$ cycles is $0.5^M$, which for sufficiently large $N$ can always contain an arbitrarily small probability mass of a *single* element, $p_n$. Therefore the number of comparison cycles to find output $n$ approaches $-\log_2(p_n)$ from above, and average number of ADC comparison cycles is lower-bounded by

$$L = -\sum_{n=0}^{2^{N-1}} p_n \cdot \log_2(p_n) \quad (1.20)$$

We established that if ADC resolution $N$ is sufficiently large, average search length of MER algorithm converges to entropy of ADC output, which according to [3] is the lower bound. The higher the number of ADC bits, the tighter the bound.

### 4. DECISION TREE EXAMPLE

To illustrate MER ADC operation, let us take example of 2 bits ADC with input $-0.5 \leq x < 3.5$, quantization step $\Delta = 1$ and the following output probability distribution:

$$p_0 = 0.125, \ p_1 = 0.125, \ p_2 = 0.25, \ p_3 = 0.5 \quad (1.21)$$

The decision trees for MER search is shown in figure 4.

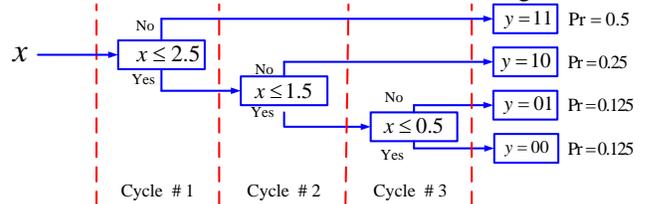

Figure 4 The SAR-ADC MER search decision tree

The conventional ADC with binary search for every sample spends 2 comparison cycles. MER ADC in the worst case, for least likely sample spends 3 cycles. However, an average number of cycles is equal to:

$$L = 1 \cdot p_3 + 2 \cdot p_2 + 3 \cdot p_1 + 3 \cdot p_0 = 1.75 \quad (1.22)$$

It illustrates the savings that MER ADC provides in the number of comparison cycles.

## 5. HARDWARE IMPLEMENTATION

The optimal threshold $\hat{y}$ for each comparison cycle may be pre-calculated according to (1.19) and stored in decision tree memory organized into ($2^N - 1$) nodes. Each node contains a comparison threshold $\hat{y}$ and two sub-nodes, for true and false comparison results respectively. Each sub-node contains a search stop flag. If this flag is true, then the sub-node contains value of ADC output. Otherwise sub-node contains the address of a new node.

Starting from node zero, the MER ADC reads threshold associated with current node and compares input signal with this threshold. Based on result of this comparison, the ADC either reads value of ADC output and stops search process, or proceeds to a new node.

The flowchart of MER search algorithm with decision tree memory implementation is shown in Figure 5. The examples of memory for MER decision tree which was shown in figure 4 is shown in Figure 6. The architecture that implements MER ADC with adaptation mechanism is shown in Figure 7.

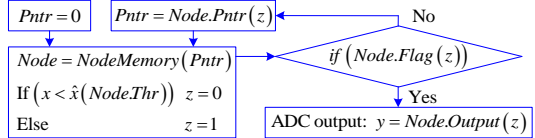

Figure 5 The flowchart of MER search algorithm with decision tree memory implementation

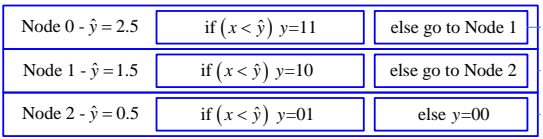

Figure 6 The MER search decision tree memory.

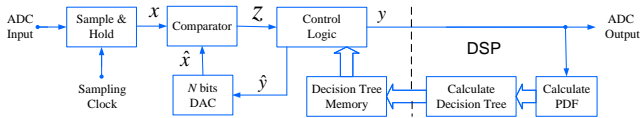

Figure 7 The adaptive MER SAR ADC

The MER ADC repeats the architecture of conventional SAR-ADC (see figure 1) with the following modifications: Control logic is equipped with decision tree memory. The Digital Signal Processor (DSP) has the ability to modify contents of the decision tree memory. At the beginning, when there is no prior information about signal statistics it stores binary decision tree and ADC operates as the conventional SAR ADC. As the conversion proceeds over successive samples, DSP collects statistics of ADC output. When statistics of ADC output settles down and input probability vector can be calculated, DSP calculates a new decision tree and stores it into ADC decision tree memory. The ADC starts operating in power saving mode.

In general, MER ADC needs input elastic buffer to absorb the sample-by-sample fluctuation of its cycle number. However, many applications that benefit from the ADC power saving do not need elastic buffer. Such applications typically require transferring a large array of analog data to digital form under a total time constraint, but do not impose a per-element time constraint. Examples of such applications are image processors and massive MIMO receivers.

## 6. SIMULATION RESULTS

For MER ADC simulation we choose signals with the following input distributions: A Gaussian distribution with peak-to-RMS ratio 10dB, and a Gaussian mixture that consists of two Gaussian distributions, one with a peak-to-RMS ratio of 10dB, the other 30dB, with likelihood of 0.1 and 0.9 respectively. First distribution is typical for OFDM signal and second one for medical imaging. Simulation results are shown in figure 7 and 8. From these figures we conclude that when ADC resolution is sufficiently high, the average number of MER ADC's comparison cycles approaches the ADC output entropy.

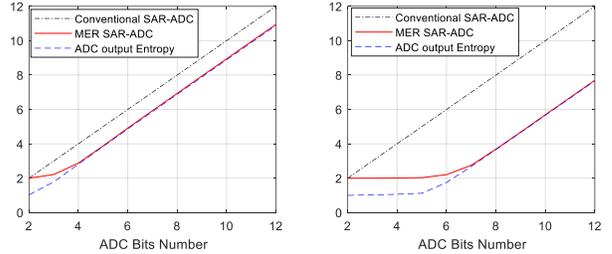

Figure 8 The average number of ADC comparison cycles for Gaussian (left) and Gaussian Mixture (right) input distributions

## 7. CONCLUSIONS

The MER search algorithm provides potentially significant reduction of SAR-ADC power consumption for signal with low entropy at the ADC output, without sacrificing ADC resolution. It may benefit power-sensitive applications such as image processing and Massive MIMO.

## 8. ACKNOWLEDGEMENT

We thank Dr. Matt Miller of Futurewei for his helpful comments.